\pdfoutput=1
\documentclass[aip,jap,reprint]{revtex4-1}

\usepackage{graphicx}
\usepackage{natbib}
\usepackage{textcomp}

\begin{document}

\title{Wireless transfer of power by a 35-GHz metamaterial split-ring resonator rectenna}

\author{Carsten Maedler}
\author{George Keiser}
\author{Adrian Yi}
\author{Jason Christopher}
\author{Mi K. Hong}
\affiliation{Department of Physics, Boston University, 590 Commonwealth Avenue, Boston, MA 02215}
\author{Alket Mertiri}
\affiliation{Division of Materials Science and Engineering, Boston University, 15 St Mary St, Brookline, MA 02446}
\author{Larry House}
\affiliation{Battelle Memorial Institute, 505 King Ave, Columbus, Ohio 43201, USA}
\author{Huseyin R. Seren}
\affiliation{Department of Mechanical Engineering, Boston University, 110 Cummington Mall, Boston, MA 02215}
\author{Xin Zhang}
\affiliation{Department of Mechanical Engineering, Boston University, 110 Cummington Mall, Boston, MA 02215}
\author{Richard Averitt}
\affiliation{Department of Physics, University of California at San Diego, 9500 Gilman Dr., La Jolla, CA 92093, USA}
\author{Pritiraj Mohanty}
\author{Shyamsunder Erramilli}
\affiliation{Department of Physics, Boston University, 590 Commonwealth Avenue, Boston, MA 02215}


\begin{abstract}
	Wireless transfer of power via high frequency microwave radiation using a miniature split ring resonator rectenna is reported. RF power is converted into DC power by integrating a rectification circuit with the split ring resonator. The near-field behavior of the rectenna is investigated with microwave radiation in the frequency range between 20-40 GHz with a maximum power level of 17 dBm. The
	observed resonance peaks match those predicted by simulation. Polarization studies show the expected maximum in signal when the electric field is polarized along the edge of the split ring resonator with the gap and minimum for perpendicular orientation. The efficiency of the rectenna is on the order of 1 \% for a frequency of 37.2 GHz. By using a cascading array of 9 split ring resonators the output power was increased by a factor of 20.
\end{abstract}


\maketitle

\section{\label{sec:introduction}Introduction}

Powering devices that are not easily accessible by hard-wired power sources or where it is impractical to regularly change batteries is an area of increasing importance including, as examples, consumer electronics, healthcare and sensor applications.\cite{pivonka2012mm, olivo2011energy, kim2012wireless, jabbar2010rf, hui2013critical, goodarzy2014feasibility, dewan2014alternative, chaimanonart2006remote, ali2005new} Delivering power remotely to these devices typically involves resonant energy transduction, followed by rectification to provide usable direct-current (DC) power. After pioneering works by Hertz and Tesla on wireless power transmission (WPT) by radio waves \cite{1234, 12345}, rectification of microwaves for highly efficient WPT was proposed in the 1960s.\cite{brown1984history} Higher operating frequencies enable shrinking of the receiving antenna. Further, the use of smaller rectifiers and direct integration with the antenna resulted in the so-called “rectenna”, leading to applications that include RF identification tags \cite{wu1996transponder} and implantable biomedical devices.\cite{matsuki1996transcutaneous, parramon1997asic} Importantly, for many applications, batteries are either too expensive or constitute health risks. Therefore, further scaling down of rectennas is crucial for use in implantable biological sensors.\citep{olivo2011energy, kotanen2012implantable}

A promising candidate for miniaturization is based on metamaterials,\cite{alavikia2014electromagnetic,fu2014broadband,hawkes2013microwave,wang2011dual} which were first realized in 1999.\cite{pendry1999magnetism, smith2000composite} Metamaterials consist of many periodically arranged single elements, which are generally much smaller than the wavelength of the electromagnetic radiation. This allows for the composite material to be described by macroscopic properties, such as electric permittivity and magnetic susceptibility. One of the most-studied metamaterial elements is a split ring resonator (SRR), which is essentially a loop with a small gap constituting an LC resonator. By varying the geometry of the SRR, its resonance frequency can be tuned independently of its size. The resonance frequency is usually obtained using numerical simulations but analytical models with excellent agreement have been proposed.\cite{sydoruk2009analytical,zhu2014circuit}

Metamaterials have been extensively studied for their artificially created exotic behavior and their fascinating applications, such as negative index of refraction,\cite{smith2000composite} electromagnetic cloaks,\cite{schurig2006metamaterial} perfect lensing \cite{pendry2000negative} and sub diffraction imaging.\cite{fang2005sub} Recently, nonlinear effects have been introduced and studied, with a focus on tunability of the metamaterial.\cite{zharov2003nonlinear} By incorporating nonlinear elements into the metamaterial new phenomena become accesible, such as bistable behavior,\cite{wang2008nonlinear} harmonic generation \cite{klein2007experiments,shadrivov2008tunable} and discrete breathers.\cite{lazarides2006discrete} Coupling of metamaterials with nonlinear elements is also necessary for their use as energy harvesters to remotely power devices with DC current.

For the purpose of this work, a rectifying diode is incorporated into the metamaterial to convert microwave radiation into usable DC power. It was predicted theoretically \cite{ramahi2012metamaterial} and has been shown experimentally \cite{hawkes2013microwave} that SRR rectennas can be used for the efficient transmission of wireless power at frequencies around 1 GHz. However, the structures used in Ref. \onlinecite{hawkes2013microwave} were too large and thus not practical for their use in powering implantable biosensors. By scaling to higher frequencies and integrating the electromagnetic functions of SRR antenna and rectifier, we demonstrate a miniaturized WPT system at a size smaller than the wavelength. For AC to DC conversion, we use a half-wave rectifier circuit with a fast Schottky diode having a low on-set voltage and a smoothing capacitor. We measured the DC voltage across a load as a function of the distance and orientation to the microwave source while varying the radiative power and microwave frequency and estimated the WPT efficiency. In order to increase the output power, we fabricated a cascading array of 9 split ring resonators.

\section{\label{sec:exp}Experimental}
\subsection{\label{sec:SRR}Rectenna Fabrication}
\begin{figure}
	\includegraphics{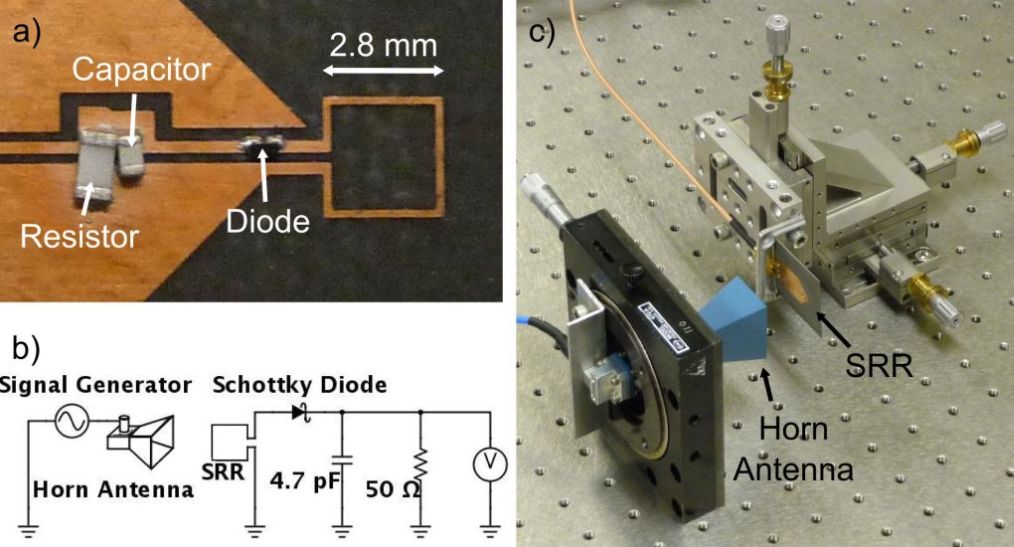}
	\caption{\label{fig:scheme}(a) Single split ring resonator with rectifying circuit. (b) Measurement setup and rectifying circuit. A signal generator supplies a signal of specific frequency and power to the horn antenna, which emits electromagnetic radiation. This couples with the SRR. The signal is then rectified and smoothed by a Schottky diode and capacitor. The DC voltage is read out by a voltmeter over a load. (c) The split ring resonator is positioned on a xyz-translational stage for movement with respect to the horn antenna which is mounted on a rotational stage. This allows for three dimensional, as well as, polarization dependent measurements.}
\end{figure}

The SRR rectenna was simulated using CST Microwave studio and designed to have a resonance frequency at 35 GHz. The SRR was fabricated on 0.79 mm thick Rogers Duroid RT5880 substrate ($\epsilon_r = 2.2$) by Advanced Circuits, Inc. The SRR material is 35 µm thick copper with the following dimensions: side length 2.8 mm, microstrip width 0.2 mm, gap size 0.2 mm. In the first part of this study all measurements were performed on single SRR elements to demonstrate the feasibility of WPT using a metamaterial rectenna. Afterwards, multiple SRR elements were combined in an array to increase the output power. In order to rectify the AC signal, we integrated a half-wave rectifying circuit with the SRR. Due to the high resonance frequency of the SRR, it is essential to use fast electrical components with low parasitic impedance. Furthermore, the signal generated by the SRR at power levels in compliance with ICNIRP regulations is small,\cite{guideline1998guidelines} hence we chose a low-barrier n-type Schottky silicon diode with small threshold voltage and small footprint (Skyworks SMS7621-060). It also features a small parasitic capacitance ($C < 0.05$ pF) and inductance ($L < 0.2$ nH). The smoothing capacitor is chosen to reduce the ripple in the output voltage (ATC 600L4R3AT200T). The output voltage $V_{out}$ is measured over a 50 Ohm load resistor using a multimeter (HP 3457A). An example of a fabricated device is shown in Figure \ref{fig:scheme}a. 

\subsection{\label{sec:meas}Measurement Setup}
For the measurements we used an Agilent signal generator (N5183A-540), which provided a maximum signal power of 19 dBm at frequencies above 30 GHz. We measured the power input into the waveguide feeding the horn antenna $P_in$ using a power meter (Agilent N1913A) with thermocouple power sensor (Agilent N8487A). The highest possible power fed into the waveguide above a frequency of 33 GHz was 17 dBm. For enhanced directivity of the generated microwave radiation we used a high gain horn antenna (Pasternack PE9850/2F-20) with a cut-off frequency of approximately 21.3 GHz. A schematic of the measurement setup is shown in Figure \ref{fig:scheme}b. Both the horn antenna and the rectenna were fixed on an optical table. The horn antenna was attached to a precision rotation stage to facilitate probing the polarity of the SRR over the full 360 degrees. The SRR was mounted on a xyz-translation stage which served two purposes. First, this allowed us to center the SRR with respect to the horn antenna, and second it enabled precise distance-dependent measurements. For all experiments the SRR was positioned with its normal parallel to the axis between horn antenna and SRR. The setup is shown in Figure \ref{fig:scheme}c.

\section{\label{sec:results}Results and Discussion}
\subsection{\label{sec:power}Power and Frequency Dependence}

\begin{figure}[!ht]
	\centering
	\includegraphics{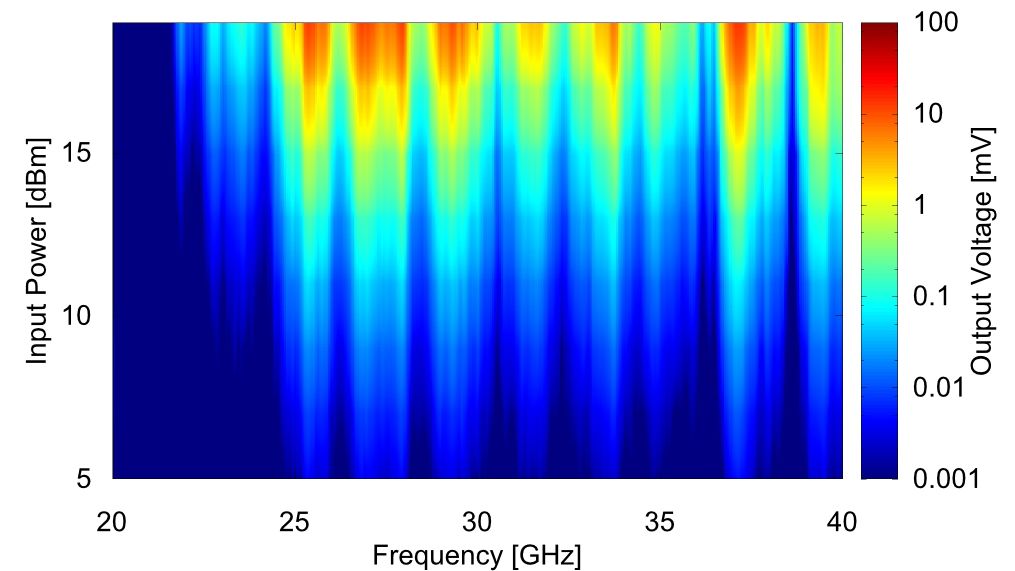}
	\caption{\label{fig:powera} The DC output voltage from a single SRR was measured for different frequencies and input powers. It is plotted logarithmically in color as shown on the right hand side of the plot. Multiple resonance frequencies are visible in the plot.}
\end{figure}

We performed frequency sweeps using a LabView program which recorded the output voltage $V_{out}$ as measured by the voltmeter for each microwave frequency and power set by the signal generator. The time between readings was set to approximately 1 second. First, we wanted to compare the simulated resonance to the actual resonance of the SRR rectenna. For this purpose we increased the frequency from 20 GHz to 40 GHz and varied the power supplied to the horn antenna from 5 to 17 dBm while recording the output voltage. For these measurements the distance and orientation between rectenna and microwave source was fixed. The data is plotted as a color plot in Figure \ref{fig:powera}, where the color corresponds to the measured output voltage. 

\begin{figure}[!ht]
	\centering
	\includegraphics{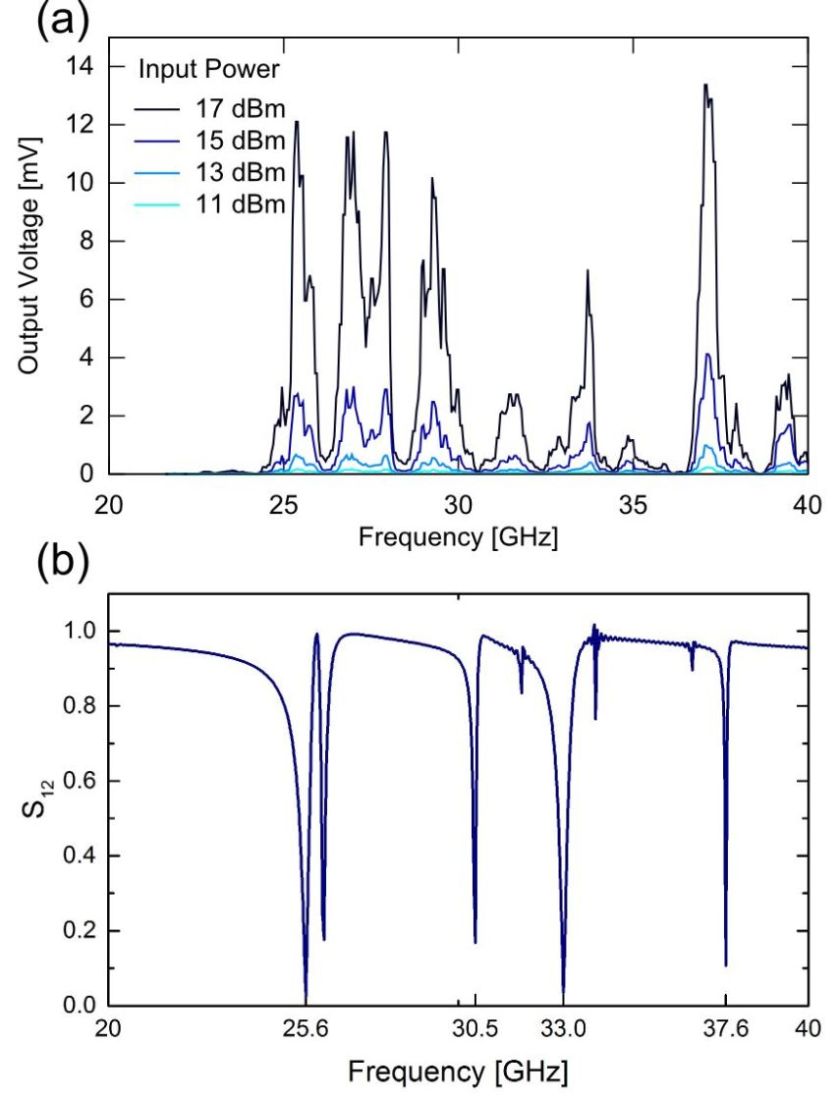}
	\caption{\label{fig:powerb} (a) The data from Figure \ref{fig:powera} is plotted in a linear plot where each line represents a different input power. The scale is linear and the lines are guides to the eye. (b) Simulated S-parameter for transmission through the SRR for comparison.}
\end{figure}

To compare the observed peaks to those predicted by simulation, the output voltage is plotted for four specific input powers in Figure \ref{fig:powerb} together with the simulated S-parameter for transmission through a SRR. The simulation was performed for linearly polarized electromagnetic radiation incident normal to the SRR structure, which was modeled by metal leads on Duroid substrate, but without rectifying elements. 

\noindent The measured peaks around 25 GHz, 30 GHz, 33 GHz, and 37 GHz match the simulated peaks. The measured peak around 28 GHz could be attributed to the additional capacitance and inductance originating in the rectifying circuit, as well as the nonlinear nature of the diode \citep{wang2008nonlinear}, and nonradiative coupling in the near-field \citep{karalis2008efficient, wavesandfields}. 

\begin{figure}[!ht]
	\centering
	\includegraphics{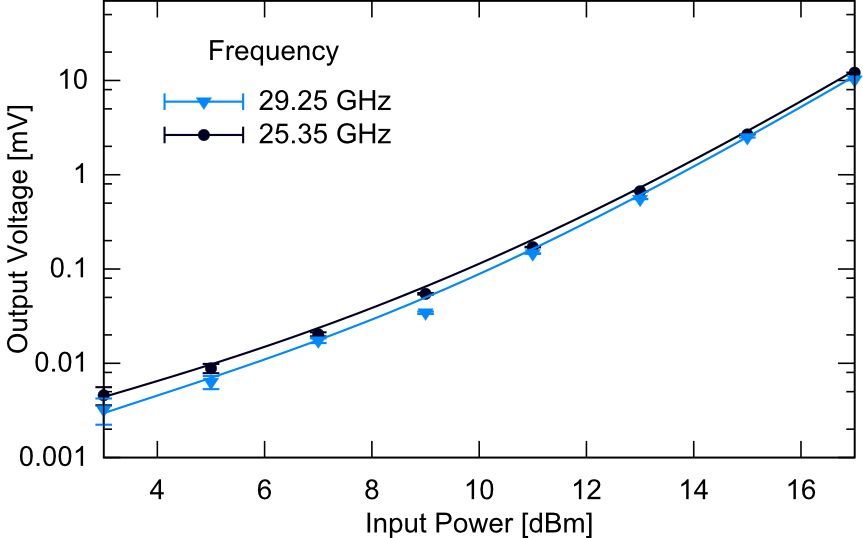}
	\caption{\label{fig:powerc}The output voltage from Figure \ref{fig:powera} is plotted for two resonance frequencies on a logarithmic scale. The error bars are determined from the noise in the measured output voltage. The lines are guides to the eye.}
\end{figure}

The output voltage is plotted with respect to the input power for two peaks on a logarithmic scale for better visualization in Figure \ref{fig:powerc}. The nonlinear relationship observed here is likely due to the nonlinear nature of the rectifying diode. For energy harvesting purposes it is advantageous to use a rectenna with larger bandwidth compared to conventional antennas. While this is not the case for WPT due to the lower efficiency at off-resonance frequencies, multi-layered resonant structures with different resonant frequencies can be used to increase bandwidth.

\subsection{\label{sec:pol}Polarization Dependence}

\begin{figure}
	\includegraphics{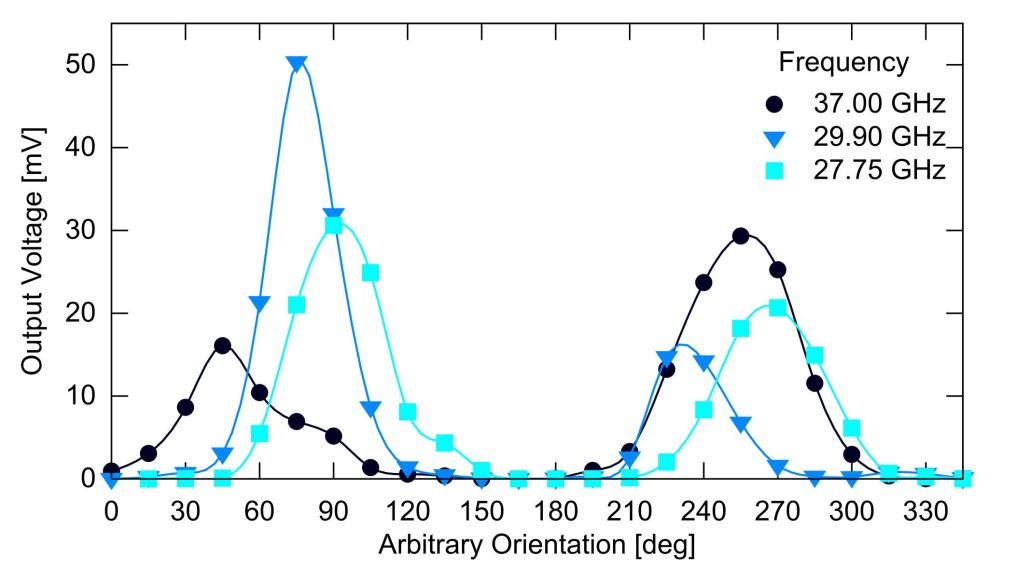}
	\caption{\label{fig:ori}A single SRR was positioned at a distance of 3 mm from the aperture of the horn antenna. The input power was set to the maximum available power from the signal generator for high frequencies. The rotation of the antenna was varied in 15 degree steps over a full circle. The DC output voltage is plotted for 3 different resonance frequencies.}
\end{figure}

The SRR is asymmetric, and we expect a strong dependence of the response on the polarization of the incident electromagnetic radiation. When the electric field is polarized parallel to the side of the SRR with the gap, a resonant LC mode is excited. In order to show this effect, we recorded the output voltage over the load while changing the angular orientation of the horn antenna with respect to the rectenna. For this experiment the distance between the SRR and horn antenna, as well as the input power, were fixed. The data is shown in the plot in Figure \ref{fig:ori}. The measured output voltage is plotted as a function of the angular orientation between SRR and horn antenna for three different resonance frequencies as identified in Figure \ref{fig:powerb}. For all three frequencies, peaks in the output voltage are visible at orientations close to where the electric field plane of the horn antenna is parallel to the side of the SRR with the gap. The output voltage drops to zero when the SRR is rotated by 90 degrees. The variations in the peak positions are likely due to nonideal polarization of the electric field and errors in the centering of the SRR. In any case, this polarization dependence shows that the resonance peaks observed in Figure \ref{fig:powerb} are indeed due to LC modes determined in part by the geometry of the SRR. 

\subsection{\label{sec:dis}Distance Dependence}

\begin{figure}
	\includegraphics{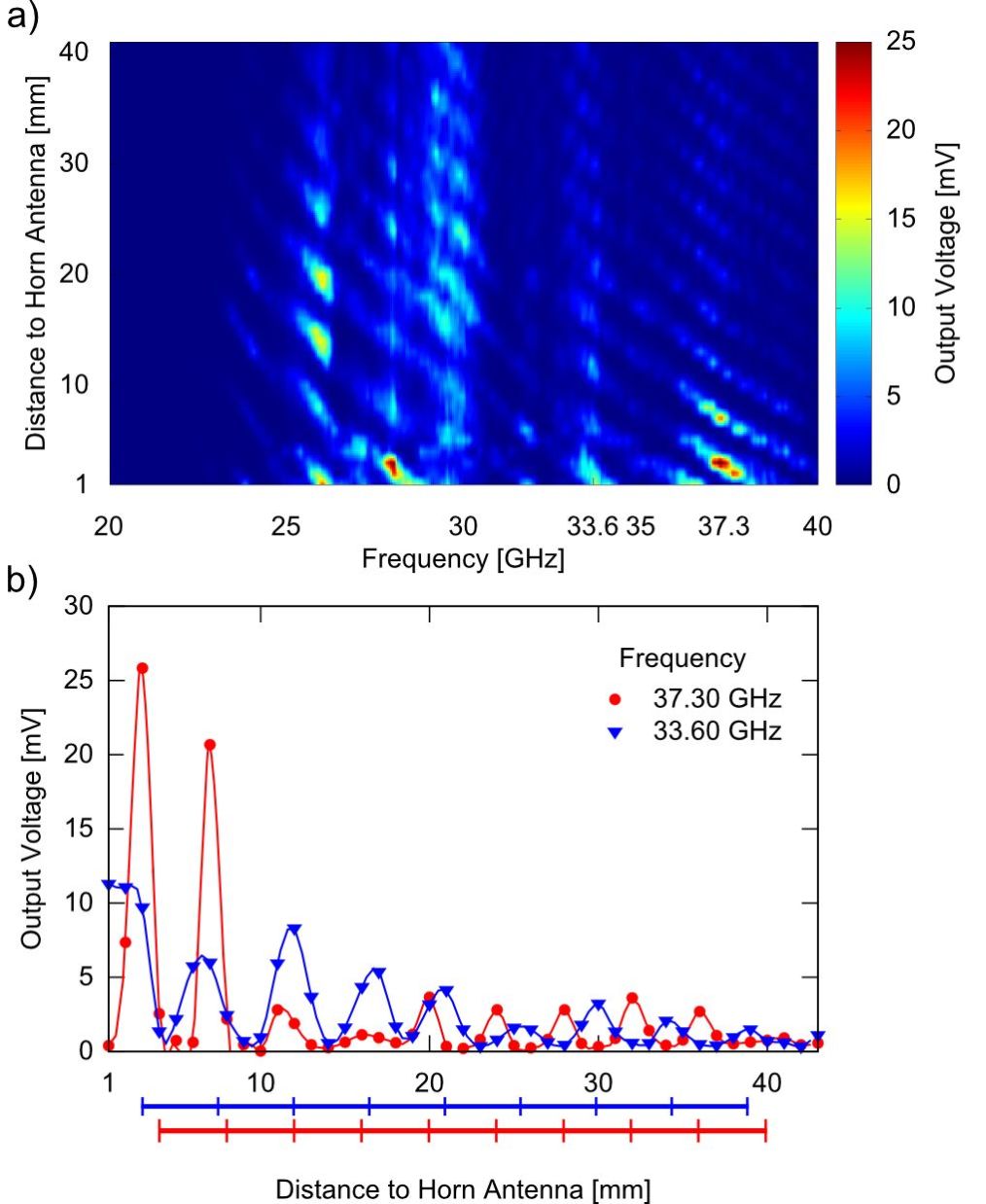}
	\caption{\label{fig:dis}(a) A single SRR was oriented for maximum coupling between the electromagnetic radiation and the SRR. The input power was set to the maximum available power from the signal generator for high frequencies. The DC output voltage is plotted for different distances in color plot as shown on the right hand side of the plot. The ripples seen in the plot indicate standing waves resulting from reflection off of the walls. (b) The output voltage from the color plot is plotted for two different resonance frequencies. The wavelength for those frequencies are 8 and 9 mm, respectively. For standing waves, maxima are, thus, expected every 4 and 4.5 mm, respectively. As a visual aid, rulers are provided below the plot with marks corresponding to those values.}
\end{figure}

Whether the rectenna is used for energy harvesting or WPT, the distance between microwave source and rectenna will be different for different applications. We studied the distance dependence of the power transfer while maintaining a constant orientation and radiation power. The data is plotted in Figure \ref{fig:dis}. The distance is measured from the SRR to the horn antenna aperture. It is evident from the data that the measured output voltage over the rectenna load decreases with distance to the horn antenna. However, a detailed analysis of the exact relationship between distance and output voltage is complicated by the fact that multiple peaks, most likely due to standing waves, appear in the plot. The peak position is periodic in distance for a single frequency and corresponds to half of the wavelength. This is illustrated in the plot of Figure \ref{fig:dis}b by two rulers, which have marks at distances of 4 and 4.5 mm, respectively, corresponding to half of the wavelength of the two graphs displayed in the plot. They align well with the observed peaks, except very close to the horn antenna. Standing waves can be utilized to increase the power transfer by mounting the SRR at a specific distance from a metal plane. In experiments with such a setup the power transfer was increased by a factor of up to ten (data not shown).

\subsection{\label{sec:eff}Efficiency Calculation}

\begin{figure}
	\includegraphics{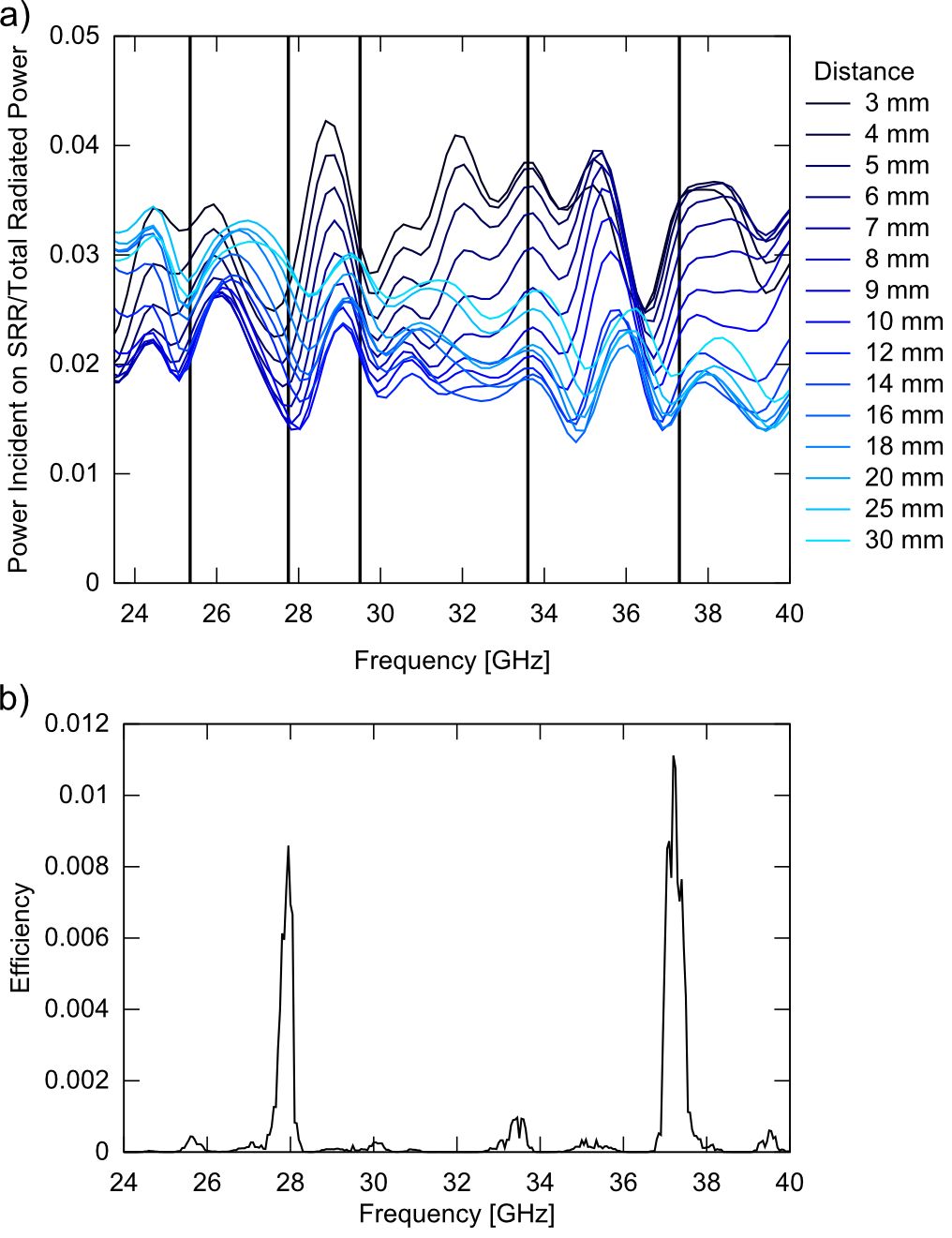}
	\caption{\label{fig:eff}(a) COMSOL simulation for the incident power on the effective area of the SRR at different distances to the horn antenna aperture and for different frequencies. (b) The measured efficiency of the rectenna as calculated by equation (3) for a distance of the SRR to the horn antenna aperture of 3 mm. The data is the same as in Figure \ref{fig:dis}. The lines are guides to the eye for both plots.}
\end{figure}

In order to calculate the efficiency of our rectenna, we need a model of the radiated power from the horn antenna. All measurements were performed at distances between the horn antenna aperture and the SRR on the order of the wavelength. Hence, the far-field approach is not applicable. However, there is no analytical formula for the power radiated from a horn antenna in the near field, requiring numerical evaluation of the equations describing the electromagnetic fields.\cite{metzger1991near} We modeled the radiation pattern using COMSOL, a finite-element, differential equation solver to calculate the power incident on the effective area $A_{eff,max}$ of the SRR. The effective area was determined according to \cite{antennatheory}

\begin{equation}
	A_{eff,max}=\frac{\lambda^2}{4\pi\times D_0},	
	\label{eq:area}
\end{equation}

\noindent where $\lambda$ is the wavelength of the microwave radiation and $D_0$ is the directivity of the SRR. In a good approximation, we can use the directivity of a small loop antenna ($D_0$ = 1.5) for the SRR.\cite{hawkes2013microwave} The power incident on the effective area $P_{incident}$ normalized by the total radiated power $P_{RF}$ is plotted in Figure \ref{fig:eff}a for different distances of the SRR to the horn antenna aperture. This ratio is henceforth denoted as $\alpha$. The available DC power $P_{out}$ from the rectenna is given by

\begin{equation}
	P_{out}=\frac{V_{out}^2}{R},		
	\label{eq:pow}
\end{equation}

\noindent where $R$ is the resistance of the load over which $V_{out}$ is measured. The efficiency $\eta$ of our rectenna can now be calculated as

\begin{equation}
	\eta=\frac{P_{out}}{P_{incident}}=\frac{P_{out}}{\alpha\times P_{in}}.		
	\label{eq:eta}
\end{equation}

The input power $P_{in}$ is measured by a power meter on the input to the horn antenna waveguide for all frequencies studied. The efficiency of the rectenna is plotted in Figure 5b for a specific distance (3 mm) of the data shown in Figure \ref{fig:dis}. It is only an approximation because the simulation did not take into account other objects or laboratory walls that were present during the measurements. The maximum efficiency observed in Figure \ref{fig:eff}b is 1 \%. This is lower compared to efficiencies achieved and predicted by other groups,\cite{hawkes2013microwave, ramahi2012metamaterial, zhu2011metamaterial} albeit for frequencies smaller by at least an order of magnitude than in our study. There are several reasons for a reduced efficiency. The main limiting factor is the rectifying diode. With increasing frequency the rectifying efficiency of the diode decreases. Furthermore, the radiation power available in this study might not have been sufficient to completely turn on the diode. The output voltage generated by the SRR was lower than the threshold voltage of the rectifying diode. A similar behavior was found by Hawkes et al. where the efficiency increased drastically with input power.\cite{hawkes2013microwave} 

\subsection{\label{sec:array}Rectenna Array}

\begin{figure}
	\includegraphics{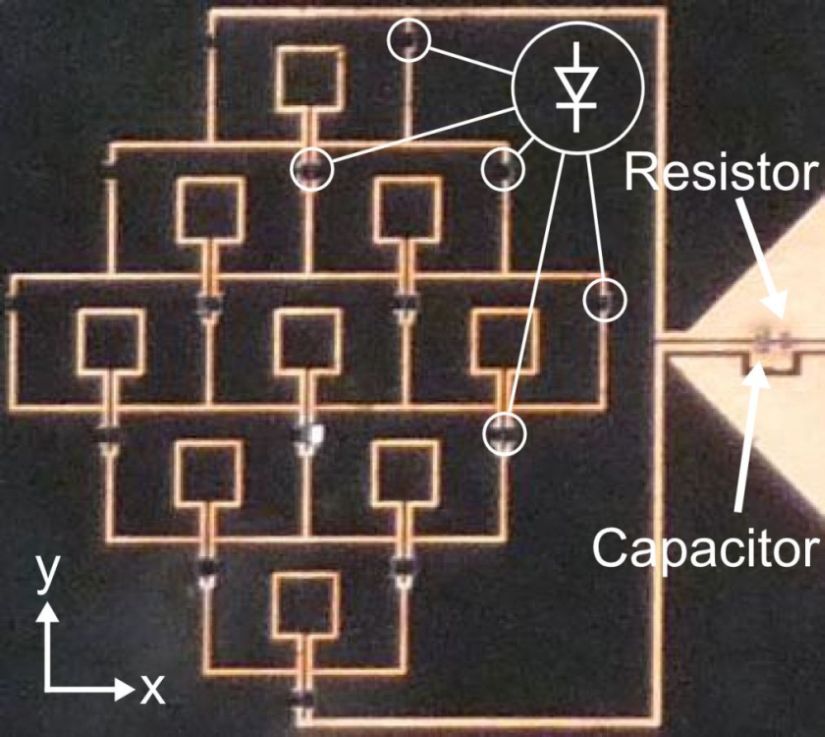}
	\caption{\label{fig:array}Rectenna array consisting of 9 split ring resonators each with 2.8 mm side length and 24 rectifying diodes, as well as a smoothing capacitor and load resistance.}
\end{figure}

The maximum measured DC output power from a single rectenna was 50 \textmu W for an input power of 50 mW, which is not sufficient for most applications. For applications that require WPT in the vicinity of humans we need a scheme to increase the output power without increasing the radiative power. Further, for energy harvesting applications the ambient RF powers collected are typically small.\cite{percy2012supplying} In order to increase the output power we devised and fabricated an array made up of nine of the single element SRRs in a cascading structure. An optical micrograph of the device is shown in Figure \ref{fig:array}. The separation of the cascading wires are 8.4 mm in x-direction and 5.4 mm in y-direction. 

\begin{figure}
	\includegraphics{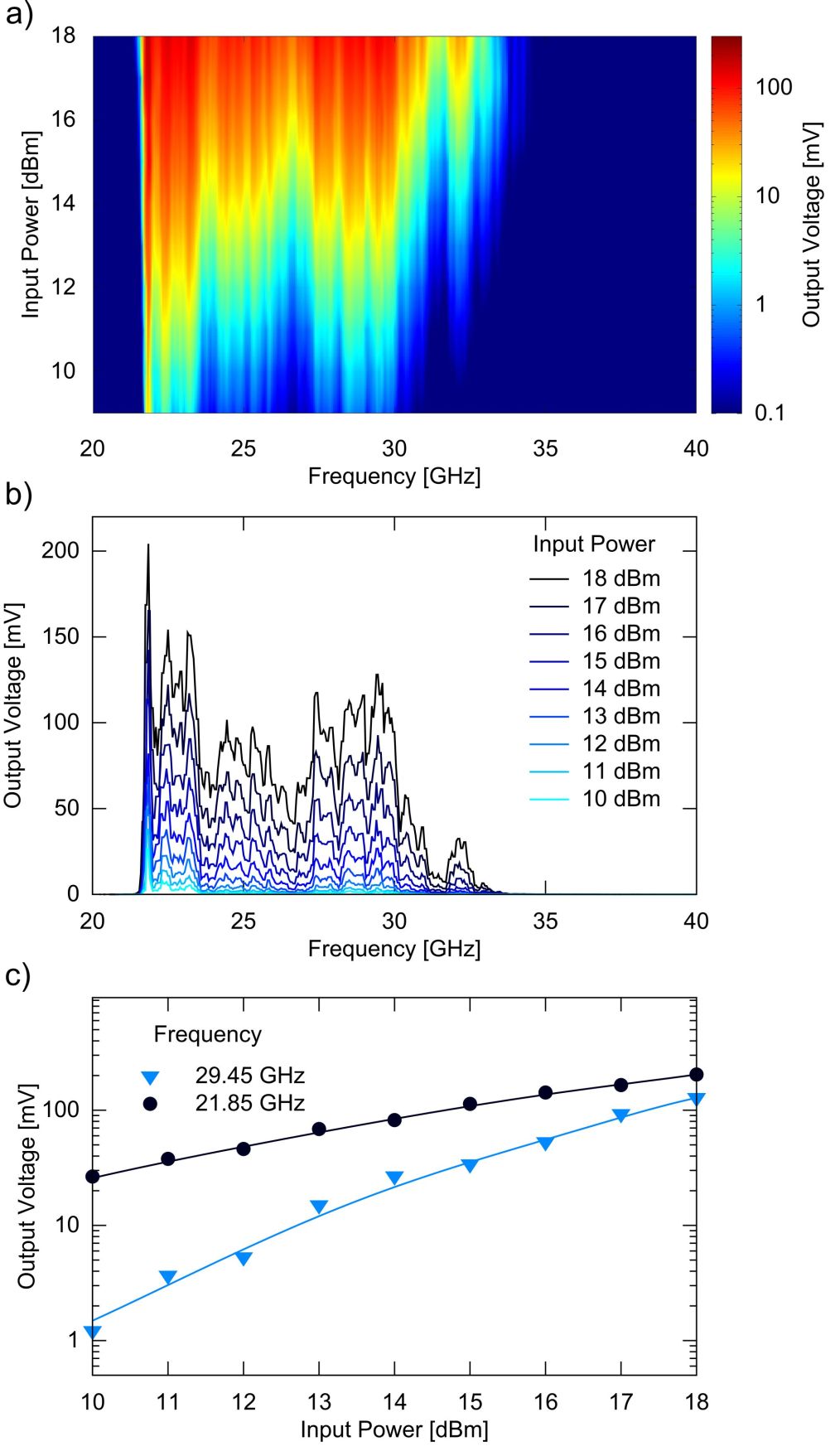}
	\caption{\label{fig:arraypow}(a) A SRR array was positioned at a distance of 3 mm from the aperture of the horn antenna. The rotation was chosen for maximum coupling of the electromagnetic radiation with the SRR. The DC output voltage was measured for different frequencies and input powers. It is plotted logarithmically in color plot as shown on the right hand side of the plot. (b) The data from the color plot is plotted in a linear plot where each line represents a different input power. The scale is linear and the lines are guides to the eye. (c) The output voltage from Figure 3a is plotted for two resonance frequencies on a logarithmic scale. The lines are guides to the eye. The error determined from the noise in the output voltage signal is on the order of 1 \textmu V and therefore too small to display in this graph.}
\end{figure}

\subsection{\label{sec:arraypow}Power and Frequency Dependence of the Array}

We first positioned the array at a close distance to the horn antenna to study the frequency-dependent response of the rectenna array (Figure \ref{fig:arraypow}a and \ref{fig:arraypow}b). Similar to the single SRR rectenna we observe more than one resonance frequency. However, the peaks are not as dominant. The measured DC voltage is higher than 40 mV over the whole frequency range from 21.7 GHz to 30 GHz for an input power of 18 dBm. This might be due to imperfections of the individual SRRs resulting in a shifted resonance frequency. The highest measured DC voltage is 204 mV at a frequency of 21.85 GHz. The signal decreases for frequencies above 30 GHz. The array was not phase matched for a single frequency because broadband operation was desired to avoid future issues with resonance matching of transmitter and receiver which is especially difficult for biological environments.\cite{fu2014broadband} The input power to DC voltage relation is plotted for two different frequencies on a logarithmic scale in Figure \ref{fig:arraypow}c. 

\subsection{\label{sec:arrayori}Polarization Dependence of the Array}

\begin{figure}
	\includegraphics{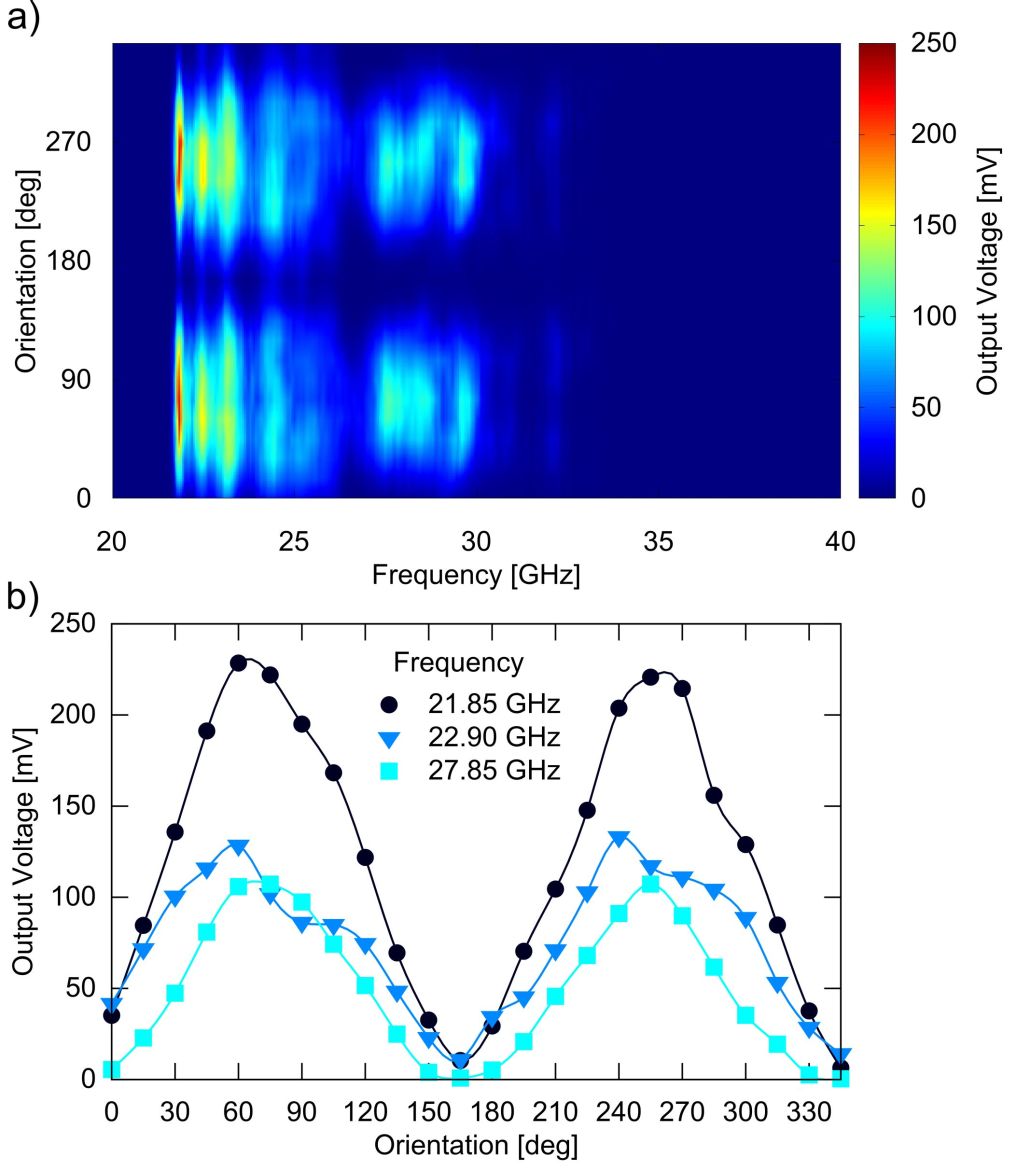}
	\caption{\label{fig:arrayori}(a) A SRR array was positioned at a distance of 3.5 mm from the aperture of the horn antenna. The input power was set to 17 dBm, the maximum available power from the signal generator for high frequencies. The rotation of the antenna was varied in 15 deg steps over a full circle. The DC output voltage is plotted in color according to the legend on the right hand side of the plot. (b) The data from the color plot is plotted in a linear plot for 3 different frequencies. The lines are guides to the eye.}
\end{figure}

The broad spectrum observed in Figure \ref{fig:arraypow} could lead one to think that the microstrips connecting the SRRs in the cascading structure act as an antenna and are at least partially responsible for receiving the microwave radiation. In order to demonstrate that the DC output voltage is due to conversion of microwave radiation by the SRR rectenna elements we performed polarization dependent measurements. For this purpose the rectenna structure was rotated for a full turn on a rotation stage around its normal axis and frequency dependent measurements were taken every 15 degrees. The DC voltage is plotted in color in Figure \ref{fig:arrayori}a for all frequencies and rotation angles. A clear polarization dependence is evident. For better analysis the DC voltage is plotted for 3 different frequencies and all angles in Figure \ref{fig:arrayori}b. For all three frequencies the output voltage is highest when the electric field plane of the horn antenna is within 15 degrees of being parallel to the side of the SRRs with the gap. At an orientation perpendicular to that configuration the output voltage drops below 11 mV. This suggests that most of the converted power originates in the SRR receiving microwave radiation.

\subsection{\label{sec:arrayeff}Output Power of the Array}

An efficiency analysis for the conversion of electromagnetic radiation power to DC power is complicated since the directivity of the array structure is not known. We can not assume a simple loop antenna as we did for the single element SRR. However, we can compare the total output power $P_{out}$ of the array to the single element. For this calculation we use the maximum measured output voltage over the load resistance $R$ of the single element ($V_{out}$ = 51.7 mV, Figure \ref{fig:ori}) and the array ($V_{out}$ = 228.6 mV, Figure \ref{fig:arrayori}b) and calculate the DC output power according to equation \ref{eq:pow}. The output power of the rectenna array is by a factor of 20 higher than that of the single rectenna. This is higher than a factor of 9 for independent elements.

\section{\label{sec:conclusion}Conclusion}

In conclusion, we fabricated and studied the behavior of a sub-wavelength-scale metamaterial rectenna by combining rectification elements directly into a split ring resonator. The measured resonance peaks match those predicted by simulation. Our device facilitates energy harvesting over a broad frequency range. The detected polarization dependence of the rectenna shows that the resonance peaks are indeed due to LC modes determined by the geometry of the SRR. The measured efficiency of our rectenna is on the order of 1 \%. This is in part due to the low input power and the high frequency used in this study. We observed standing wave patterns due to reflection of the microwave radiation off the walls in the building, which might serve as a power transmission enhancement scheme by incorporating a metal plane at a specific distance behind the rectenna. Preliminary results show an enhancement in efficiency by up to a factor of 10. In order to improve the efficiency further the rectifying circuit should be optimized.\cite{hagerty2004recycling, curty2005model} In an effort to increase the output power we fabricated and studied an array of nine of single SRRs. Here, we observed a broadband wireless power transfer from frequencies between 21.7 GHz and 30 GHz with the highest DC output power at 21.85 GHz. Compared to the single element SRR rectenna we achieved a DC power output increase of a factor of 20. The successful demonstration of WPT using a metamaterial rectenna structure should stimulate further research on possible applications of those structures, such as wireless powering of biological implants and sensors, or remote sensor networks.

\begin{acknowledgments}
	We thank Thomas Dudley for help and assistance. This work was supported by Battelle Memorial Institute and the National Science Foundation (PFI-AIR 1237848). HS and XZ would like to acknowledge NSF ECCS-1309835. 
\end{acknowledgments}

\bibliography{apssamp}

\end{document}